\pdfoutput=0
\documentclass[11pt]{article}
\usepackage{axodraw}
\usepackage{epsfig}
\usepackage{amsfonts}
\usepackage{amsmath}
\usepackage{bm,bbm}
\usepackage{cite}
\usepackage{hyperref}
\allowdisplaybreaks[1]

\input pix.sty

\newcommand{\la}[1]{\label{#1}}
\newcommand{\be}{\begin{equation}}
\newcommand{\ee}{\end{equation}}
\newcommand{\ba}{\begin{eqnarray}}
\newcommand{\ea}{\end{eqnarray}}
\newcommand{\rmi}[1]{{\mbox{\scriptsize #1}}}
\newcommand{\fig}{Fig.~}

\newcommand{\eq}{Eq.~}
\newcommand{\eqs}{Eqs.~}
\newcommand{\se}{Sec.~}

\newcommand{\nr}[1]{(\ref{#1})}

\newcommand{\nn}{\nonumber \\}

\renewcommand{\vec}[1]{{\bf #1}}

\renewcommand{\eq}{eq.~}
\renewcommand{\eqs}{eqs.~}
\renewcommand{\se}{sec.~}

\renewcommand{\fig}{fig.~}

\newcommand{\Nc}{N_{\rm c}}

\newcommand{\now}{{\mbox{\tiny\rm{0}}}}
\newcommand{\fnow}{f_\now}
\newcommand{\Tnow}{T_\now}
\newcommand{\tnow}{t_\now}

\newcommand{\Tmax}{T_{\rm max}}

\newcommand{\rmO}{{\mathcal{O}}}

\def\lsi{\raise0.3ex\hbox{$<$\kern-0.75em\raise-1.1ex\hbox{$\sim$}}}
\def\gsi{\raise0.3ex\hbox{$>$\kern-0.75em\raise-1.1ex\hbox{$\sim$}}}

\newcommand{\nF}{f_\rmii{F}}
\newcommand{\nB}{f_\rmii{B}}
\newcommand{\rmii}[1]{{\mbox{\tiny\rm{#1}}}}
\newcommand{\rmiii}[1]{{\mbox{\tiny{$\scriptstyle{\rm#1}$}}}}

\newcommand{\Tint}[1]{{\hbox{$\sum$}\!\!\!\!\!\!\!\int\,}_{\!\!\!\!\raise-0.9ex\hbox{$\scriptstyle{#1}$}}}
\newcommand{\Tinti}[1]{{{\Sigma}\!\!\!\!\raise0.3ex\hbox{$\int$}_\rmii{${#1}$}}}
\newcommand{\bi}{\begin{itemize}}
\newcommand{\ei}{\end{itemize}}
\newcommand{\hide}[1]{ }
\newcommand{\bsl}[1]{\,\slash\!\!\!\!{#1}\,}

\newcommand{\deltabar}{\raise-0.02em\hbox{$\bar{}$}\hspace*{-0.8mm}{\delta}}
\newcommand{\ddeltabar}{\raise-0.18em\hbox{$\bar{}$}\hspace*{-0.8mm}{\delta}}
\renewcommand{\P}{\mathcal{P}}

\newcommand{\K}{\mathcal{K}}

\newcommand{\mpl}{m_\rmii{pl}} 
\newcommand{\lnf}{l^{ }_\rmi{1f}}
\newcommand{\lnb}{l^{ }_\rmi{1b}}
\newcommand{\lf}[1]{l^{ }_\rmi{#1f}}
\newcommand{\lb}[1]{l^{ }_\rmi{#1b}}
\newcommand{\qm}{q_{-}}
\newcommand{\qp}{q_{+}}
\newcommand{\gw}{\rmi{gw}}
\newcommand{\igw}{\rmii{gw}}
\newcommand{\bbn}{\rmi{bbn}}
\def\TAsc(#1,#2)(#3,#4,#5)%
{\SetWidth{2.0}\CArc(#1,#2)(#3,#4,#5)\SetWidth{1.0}}

\def\TAgl(#1,#2)(#3,#4,#5){\SetWidth{2.0}\PhotonArc(#1,#2)(#3,#4,#5){\Lwidth}%
{6.283 #3 mul 360 div #4 #5 sub #4 #5 sub mul sqrt mul Tdensity mul}%
\SetWidth{1.0}}
\def\TLgl(#1,#2)(#3,#4){\SetWidth{2.0}\Photon(#1,#2)(#3,#4){\Lwidth}
{#1 #3 sub #1 #3 sub mul #2 #4 sub #2 #4 sub mul add sqrt Tdensity mul}%
\SetWidth{1.0}}

\renewcommand{\picc}[1]{\;\parbox[c]{60pt}{\begin{picture}(60,30)(0,-10)
\SetWidth{1.0}\SetScale{1.3} #1 \end{picture}}\; }

\def\Lwidth{2}

\newcommand{\picw}[1]{\;\parbox[c]{45pt}{\begin{picture}(70,30)(-10,5)
\SetWidth{1.0}\SetScale{0.9} #1 \end{picture}}\; }
\newcommand{\picx}[1]{\;\parbox[c]{45pt}{\begin{picture}(70,30)(-10,5)
\SetWidth{1.0}\SetScale{0.8} #1 \end{picture}}\; }

\def\AmplAscalar{\picw{%
 \Lsc(10,40)(29,21.5)%
 \Lsc(10,0)(29,18.5)%
 \Lhh(30.5,20)(50.5,39)%
 \Lhh(29,21.5)(49,40.5)%
 \Lhh(29,18.5)(49,-0.5)%
 \Lhh(30.5,20)(50.5,1)%
}}
\def\AmplBscalar{\picw{%
 \Lsc(10,40)(30,35)%
 \Lsc(10,0)(30,5)%
 \Lsc(30.5,7)(30.5,33)%
 \Lhh(30,35)(50,40)%
 \Lhh(30.5,33)(50.5,38)%
 \Lhh(30.5,7)(50.5,2)%
 \Lhh(30,5)(50,0)%
}}
\def\AmplCscalar{\picx{%
 \Lsc(10,40)(31,35)%
 \Lsc(10,0)(31,5)%
 \Lsc(29.2,6.2)(29.2,33.8)%
 \Lhh(31,35)(56,0)%
 \Lhh(29.2,33.8)(54.2,-1.2)%
 \CCirc(40,20){3}{White}{White}
 \Lhh(29.2,6.2)(54.2,41.2)%
 \Lhh(31,5)(56,40)%
}}
\def\AmplDscalar{\picx{%
 \Lsc(0,40)(19,21)%
 \Lsc(0,0)(19,19)%
 \Lhh(19,21)(41,21)%
 \Lhh(19,19)(41,19)%
 \Lhh(41,21)(61,41)%
 \Lhh(42.5,19.5)(62.5,39.5)%
 \Lhh(42.5,20.5)(62.5,0.5)%
 \Lhh(41,19)(61,-1)%
}}
\def\AmplAfermion{\picw{%
 \Lqq(10,40)(29,21.5)%
 \Lqq(10,0)(29,18.5)%
 \Lhh(30.5,20)(50.5,39)%
 \Lhh(29,21.5)(49,40.5)%
 \Lhh(29,18.5)(49,-0.5)%
 \Lhh(30.5,20)(50.5,1)%
}}
\def\AmplBfermion{\picw{%
 \Lqq(10,40)(30,35)%
 \Lqq(10,0)(30,5)%
 \Lqq(30.5,7)(30.5,33)%
 \Lhh(30,35)(50,40)%
 \Lhh(30.5,33)(50.5,38)%
 \Lhh(30.5,7)(50.5,2)%
 \Lhh(30,5)(50,0)%
}}
\def\AmplCfermion{\picx{%
 \Lqq(10,40)(31,35)%
 \Lqq(10,0)(31,5)%
 \Lqq(29.2,6.2)(29.2,33.8)%
 \Lhh(31,35)(56,0)%
 \Lhh(29.2,33.8)(54.2,-1.2)%
 \CCirc(40,20){3}{White}{White}
 \Lhh(29.2,6.2)(54.2,41.2)%
 \Lhh(31,5)(56,40)%
}}
\def\AmplDfermion{\picx{%
 \Lqq(0,40)(19,21)%
 \Lqq(0,0)(19,19)%
 \Lhh(19,21)(41,21)%
 \Lhh(19,19)(41,19)%
 \Lhh(41,21)(61,41)%
 \Lhh(42.5,19.5)(62.5,39.5)%
 \Lhh(42.5,20.5)(62.5,0.5)%
 \Lhh(41,19)(61,-1)%
}}
\def\AmplAgauge{\picw{%
 \Lgl(10,40)(29,21.5)%
 \Lgl(10,0)(29,18.5)%
 \Lhh(30.5,20)(50.5,39)%
 \Lhh(29,21.5)(49,40.5)%
 \Lhh(29,18.5)(49,-0.5)%
 \Lhh(30.5,20)(50.5,1)%
}}
\def\AmplBgauge{\picw{%
 \Lgl(10,40)(30,35)%
 \Lgl(10,0)(30,5)%
 \Lgl(30.5,7)(30.5,33)%
 \Lhh(30,35)(50,40)%
 \Lhh(30.5,33)(50.5,38)%
 \Lhh(30.5,7)(50.5,2)%
 \Lhh(30,5)(50,0)%
}}
\def\AmplCgauge{\picx{%
 \Lgl(10,40)(31,35)%
 \Lgl(10,0)(31,5)%
 \Lgl(29.2,6.4)(29.2,33.8)%
 \Lhh(31,35)(56,0)%
 \Lhh(29.2,33.8)(54.2,-1.2)%
 \CCirc(40,20){3}{White}{White}
 \Lhh(29.2,6.2)(54.2,41.2)%
 \Lhh(31,5)(56,40)%
}}
\def\AmplDgauge{\picx{%
 \Lgl(0,40)(19,21)%
 \Lgl(0,0)(19,19)%
 \Lhh(19,21)(41,21)%
 \Lhh(19,19)(41,19)%
 \Lhh(41,21)(61,41)%
 \Lhh(42.5,19.5)(62.5,39.5)%
 \Lhh(42.5,20.5)(62.5,0.5)%
 \Lhh(41,19)(61,-1)%
}}
\makeatletter \@addtoreset{equation}{section} \makeatother
\renewcommand{\theequation}{\arabic{section}.\arabic{equation}}
\makeatletter
\renewcommand\section{\@startsection {section}{1}{\z@}%
                                   {-5.5ex \@plus -1ex \@minus -.2ex}
                                   {2.3ex \@plus.2ex}%
                                   {\normalfont\large\bfseries}}
\renewcommand\subsection{\@startsection{subsection}{2}{\z@}%
                                     {-3.25ex\@plus -1ex \@minus -.2ex}%
                                     {1.5ex \@plus .2ex}%
                                     {\normalfont\normalsize\bfseries}}
\renewcommand\thesection {\@arabic\c@section}
\renewcommand\thesubsection   {\thesection.\@arabic\c@subsection}
\renewcommand{\@seccntformat}[1]{%
\csname the#1\endcsname.\hspace{1.0em}}
\makeatother

\begin{document}

\flushbottom

\begin{titlepage}

\begin{flushright}
RIKEN-iTHEMS-Report-24 \\
CERN-TH-2023-216 \\   
March 2024
\end{flushright}

\vfill

\begin{centering}

{\Large{\bf
    Double-graviton production from Standard Model plasma
}} 

\vspace{0.8cm}

J.~Ghiglieri$^\rmi{\hspace*{0.2mm}a}_{ }$, 
M.~Laine$^\rmi{\hspace*{0.2mm}b}_{ }$,
J.~Sch\"utte-Engel$^\rmi{\hspace*{0.2mm}c,d}_{ }$, 
E.~Speranza$^\rmi{\hspace*{0.2mm}e}_{ }$ 

\vspace{0.6cm}

${}^\rmi{a}_{ }${\em
SUBATECH,
Nantes Universit\'e, 
IMT Atlantique, 
IN2P3/CNRS, \\ 
4 rue Alfred Kastler, 
La Chantrerie BP 20722, 
44307 Nantes, France \\}

\vspace*{0.3cm}

${}^\rmi{b}_{ }${\em
AEC, 
Institute for Theoretical Physics, 
University of Bern, \\ 
Sidlerstrasse 5, CH-3012 Bern, Switzerland \\}

\vspace*{0.3cm}

${}^\rmi{c}_{ }${\em
Department of Physics, 
University of California, Berkeley, CA 94720, USA 
\\}

\vspace*{0.3cm}

${}^\rmi{d}_{ }${\em
RIKEN iTHEMS, Wako, Saitama 351-0198, Japan 
\\}

\vspace*{0.3cm}

${}^\rmi{e}_{ }${\em
Theoretical Physics Department, CERN, \\ 
CH-1211 Geneva 23, Switzerland \\}

\vspace*{0.6cm}

{\em 
Emails: 
jacopo.ghiglieri@subatech.in2p3.fr, 
laine@itp.unibe.ch, \\ 
janschue@berkeley.edu, 
enrico.speranza@cern.ch
}

\vspace*{0.8cm}

\mbox{\bf Abstract}
 
\end{centering}

\vspace*{0.3cm}
 
\noindent
The thermal plasma filling the early universe generated a stochastic
gravitational wave background that peaks in the microwave frequency
range today. If the graviton production rate is expressed as a series in a 
fine-structure constant, $\alpha$, and the temperature over the Planck mass,
$T^2_{ } / \mpl^2$, then the lowest-order contributions come from single
($\sim \alpha T^2_{ }/\mpl^2$) and double ($\sim T^4_{ }/\mpl^4$)
graviton production via $2\to 2$ scatterings. 
We show that in the Standard Model, 
single-graviton production dominates if the maximal 
temperature is smaller than 
$4\times 10^{18}_{ }$~GeV. This justifies previous calculations 
which relied solely on single-graviton production.
We mention Beyond the Standard Model scenarios in which the single and 
double-graviton contributions could be of comparable magnitudes. 
Finally, we elaborate on what these results imply for the range of 
applicability of General Relativity as an effective theory. 

\vfill


\end{titlepage}

\tableofcontents

%
\section{Introduction}
\la{se:intro}

The past years have seen an explosion of interest 
in gravitational wave astronomy. 
With hopes for an upcoming observation of a primordial background
through pulsar timing arrays~\cite{nhz1,nhz2,nhz3,nhz4},
the construction of the LISA observatory under way, and 
the planning of DECIGO as well as 
the Einstein Telescope striding forward,  
theoretical efforts at identifying all conceivable 
cosmological sources are well motivated.
In fact such efforts should cover not only the sub-kHz frequencies 
relevant for large interferometers, 
but extend up to much higher frequencies, 
100\hspace*{0.5mm}GHz and above, 
for which future experimental concepts 
are being actively developed~\cite{uhf_rev}.

If we restrict ourselves to the Standard Model of particle physics, 
perhaps minimally extended to incorporate the existence 
of neutrino masses, there is however only one primordial gravitational
wave component that is guaranteed to be present.\footnote{%
 Many other sources have been analyzed, 
 notably inflation~\cite{gw_infl}, 
 the dynamics of reheating~\cite{gw_preheat1,kl}, 
 and a plethora of post-reheating phenomena~\cite{cf,rw}, 
 however they rely on Beyond the Standard Model (BSM) physics. 
 }
It is the gravitational waves produced 
by microscopic collisions~\cite{sw}
taking place in a Standard Model 
plasma~\cite{gravity_qualitative,gravity_lo}. 
This background 
is {\em not} a perfect analogue of the 
cosmic microwave background (CMB), 
since gravitons were most likely
never in thermal equilibrium,\footnote{%
 Exotic scenarios in which this could have been the case have also
 been proposed, cf.,\ e.g.,\ refs.~\cite{ba,vl}.
 } 
however its shape ($\sim$ Planckian)
and peak frequency range 
($\fnow^{ } |^{ }_\rmi{peak}\sim 100$\hspace*{0.3mm}GHz) 
are similar to those of the CMB. The reason for the similarity
is that the graviton energies 
reflect the thermal energy distributions of the 
plasma particles from which they are produced. 

A key property of the thermal gravitational wave background is 
that while its shape can be computed, its amplitude is 
not known. The reason is that gravitons do not equilibrate;
they are just continuously being produced. The production rate is 
proportional to the strength of the gravitational interaction, 
$\sim 1/\mpl^2$, where $\mpl^{ }$ is the Planck mass, as well
as to the strength of Standard Model forces, $\sim \alpha$, 
where $\alpha$ is a fine-structure constant. Integrating
over a Hubble time, $H^{-1}_{ } \sim \mpl^{ }/T^2 $, and normalizing
to the energy density of radiation, the fractional energy 
density in gravitational waves then scales as 
$\Omega^{ }_\gw \sim \alpha \Tmax^{ } / \mpl^{ }$, 
where $\Tmax^{ }$ is the maximal temperature of the 
radiation-dominated phase in our universe.  

Now, if we explore large values of $\Tmax^{ }$, it must be
asked whether terms suppressed by higher powers of $\Tmax^{ }/\mpl^{ }$
also play a role. A concrete way to probe this question
was identified in ref.~\cite{gss}. The rate 
$\sim \alpha/\mpl^2$ mentioned above involves the production of 
a single graviton in association with a Standard Model particle. 
But in addition, there are processes in which two gravitons 
are produced, without any Standard Model particles. 
The rate of the latter processes is $\sim 1/\mpl^4$.
To complete the dimensions, this rate must come with 
a power of $\Tmax^{ }$. 
Therefore, at high $\Tmax^{ }$, the 
double-graviton rate overtakes the single-graviton rate. 
Then we also need to ask whether the notion of using 
General Relativity as an effective theory, reliant on the convergence
of an expansion in $\Tmax^2/\mpl^2$, continues to apply. 

The purpose of the current study is to promote the computation 
of ref.~\cite{gss}, carried out in scalar field theory, to the 
full Standard Model. This requires including fermion-antifermion
pairs and gauge boson pairs as additional initial states. Summing over 
all processes, we provide a quantitative criterion 
for when single-graviton production dominates. 

Our presentation is organized as follows. 
The main steps of the computation, as well as a difference 
to matrix elements squared that can be found in the literature, 
are discussed in \se\ref{se:outline}. Our results
are illustrated numerically in \se\ref{se:plots}, 
and the main conclusions are formulated in \se\ref{se:concl}. 
Some technical details related to evaluating thermal 
averages of pair production cross sections have been  
relegated to appendix~A.

%
\section{Steps of the computation}
\la{se:outline}

We consider a moment after inflation at which the plasma filling
the universe has equilibrated to an average temperature $T \ll \mpl^{ }$. 
Furthermore, 
the plasma contribution to the overall energy density is assumed
to dominate over that from non-equilibrated fields, such as
the inflaton. Then the Hubble rate, $H$, is much smaller
than typical thermal momenta, 
$H \sim \sqrt{g^{ }_*}\,T^2_{ } / \mpl^{ } \ll \pi T$, 
where $g^{ }_* \sim 10^2_{ }$ 
is the number of relativistic degrees of freedom. 
Consequently, thermal wavelengths
are well within the horizon, $(\pi T)^{-1}_{ } \ll H^{-1}_{ }$. We consider
the production of gravitational waves (GW's) from the scatterings of 
particles with such wavelengths. 
As will become apparent, the GW spectrum reflects that of
the scatterers, 
so that in terms of the physical momentum $k$, 
it is peaked at around $k | ^{ }_\rmi{peak} \sim \pi T$.

Given that the physics we are interested in takes place within
the horizon, the space is locally flat, and we can employ local 
Minkowskian coordinates for the computation. The background metric
is written as 
$
 g^{ }_{\mu\nu} = \eta^{ }_{\mu\nu} + \kappa\, h^{ }_{\mu\nu}
$, 
where 
$
 \eta^{ }_{\mu\nu} = 
 \mathop{\mbox{diag}}
$($+$$-$$-$$-$)
is the Minkowski metric and 
$h^{ }_{\mu\nu}$ denotes the gravitational perturbation. 
The effective coupling of gravitational interactions is defined as 
\be
 \kappa^2_{ } \; \equiv \; 32\pi G \; \equiv \; \frac{32\pi}{\mpl^2}
 \;, \la{kappa_def}
\ee
where $G$ is Newton's constant, and 
$\mpl^{ } = 1.22091 \times 10^{19}_{ } \; \mbox{GeV}$
is the Planck mass.

When the Einstein-Hilbert action is expanded as a series 
in $\kappa$, vertices are generated 
for the interactions of $h^{ }_{\mu\nu}$
with itself and with other fields 
(a pedagogic review can be found in ref.~\cite{choi}).
Adding a graviton leg to a process leads to a suppression
of the amplitude by~$\kappa \sim 1/\mpl^{ }$, and to a suppression
of the corresponding amplitude squared by $\kappa^2_{ }$.
But we can also add a normal Standard Model vertex to any
given process, which leads to the suppression of the rate
by $\sim \alpha$, where $\alpha$ is a fine-structure constant. 
Therefore, the production rate can be viewed as a double series, 
in $\alpha$ and $1/\mpl^2$. In addition, we may
recall that $2\to 1$ processes are kinematically forbidden
for a massless final state. If we approximate all 
particles as massless, the leading processes are
then $2\to 2$ amplitudes. If they involve a Standard Model
vertex, their rate is $\sim \alpha/\mpl^2$; if they only
contain gravitational vertices, their rate is $\sim 1/\mpl^4$.
To these leading terms, additional vertices of either type
can be added, leading to further suppression by $\alpha$
or $1/\mpl^2$ (e.g.\ triple-graviton production
at $\sim 1/\mpl^6$, etc). 

\vspace*{3mm}

\begin{figure}[t]
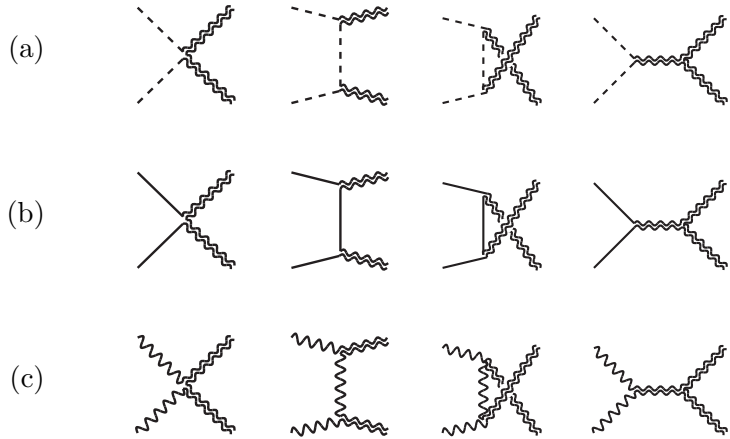


\begin{eqnarray*}
  \mbox{(a)}
  \hspace*{-2.5mm} 
  && 
  \hspace*{0.0mm}
  \AmplAscalar  
  \hspace*{2.5mm}
  \AmplBscalar 
  \hspace*{2.5mm}
  \AmplCscalar
  \hspace*{5.0mm}
  \AmplDscalar
  \\[10mm] 
  \mbox{(b)}
  \hspace*{-2.5mm} 
  && 
  \hspace*{0.0mm}
  \AmplAfermion  
  \hspace*{2.5mm}
  \AmplBfermion 
  \hspace*{2.5mm}
  \AmplCfermion
  \hspace*{5.0mm}
  \AmplDfermion
  \\[10mm] 
  \mbox{(c)}
  \hspace*{-2.5mm} 
  && 
  \hspace*{0.0mm}
  \AmplAgauge  
  \hspace*{2.5mm}
  \AmplBgauge 
  \hspace*{2.5mm}
  \AmplCgauge
  \hspace*{5.0mm}
  \AmplDgauge
  \hspace*{15mm} 
\end{eqnarray*}

\vspace*{2mm}

\caption[a]{\small
  Diagrams leading to the pair production of gravitons
  (double wiggly lines) from 
  (a)~scalars (dashed lines); 
  (b)~fermions (solid lines);  
  (c)~gauge fields (wavy lines). 
  We have omitted arrows from scalars and fermions, 
  understanding that both particle and antiparticle states
  are to be included. 
}

\la{fig:graphs}
\end{figure}

The practical computations can be streamlined by going over to the 
De Donder gauge, 
$
 \partial^{ }_\alpha h^\alpha_\mu = 
 \frac{1}{2} \partial^{ }_\mu h^\alpha_\alpha
$. 
This also renders the quadratic part of the gravitational action invertible, 
so that an internal graviton propagator can be found.
On the external graviton lines, 
which are on-shell, i.e.\ with four-momentum 
$\K = (k,\vec{k})$ with $k \equiv |\vec{k}|$, 
the sum over polarizations $\lambda$ yields the projector
\be
 \sum_{\lambda = \{ \times,+ \} } 
                \epsilon^\lambda_{\alpha\beta}(\vec{k})\, 
                \epsilon^{\lambda*}_{\mu\nu}(\vec{k})
 \; \equiv \;  
 \mathbbm{L}^{ }_{\alpha\beta;\mu\nu}
 \; = \; 
  \frac{1}{2}
  \Bigl(  
      \mathbbm{K}^\rmii{T}_{\alpha\mu}\mathbbm{K}^\rmii{T}_{\beta\nu} 
   +  \mathbbm{K}^\rmii{T}_{\alpha\nu}\mathbbm{K}^\rmii{T}_{\beta\mu}
   -  \mathbbm{K}^\rmii{T}_{\alpha\beta}\mathbbm{K}^\rmii{T}_{\mu\nu}
  \Bigr)
  \;, 
\ee  
which is traceless in $\alpha\beta$ and $\mu\nu$, 
transverse with respect to the graviton four-momentum $\K$, 
and projects onto two 
physical states in four dimensions, 
i.e.\ 
$
 {\mathbbm{L}^{ }_{\alpha\beta;}}^{\alpha\beta}_{} = 2
$. 
The tensor $\mathbbm{K}^\rmii{T}_{ }$ can be chosen as
\be
 \mathbbm{K}^\rmii{T}_{\mu\nu}
 \; \stackrel{\rm }{\equiv} \; 
 \eta^{ }_{\mu i}\eta^{ }_{\nu j} 
 \biggl( \delta^{ij}_{ } -  \frac{k^i_{ }k^j_{ }}{k^2_{ }} \biggr)
 \; = \; 
 -\, \eta^{ }_{\mu\nu} 
 + \frac{\K^{ }_\mu \bar\K^{ }_\nu + \K^{ }_\nu \bar\K^{ }_\mu}
        {\K \cdot \bar\K}
 \;, 
 \la{K_T} 
\ee  
which is 
transverse with respect to the four-momenta 
$\K$ and 
$\bar\K \equiv (k,-\vec{k})$, 
and also with respect to 
the three-momentum $\vec{k}$ (we work in the plasma rest frame). 

With the Feynman rules at hand,
the first step is to determine the matrix elements squared for the 
production processes, shown 
in \fig\ref{fig:graphs}. The four-momenta of the Standard Model 
particles are denoted by $\P^{ }_{1}$, $\P^{ }_2$ and those of the gravitons
by $\K^{ }_{1}$, $\K^{ }_2$, 
with $\P^{ }_1 + \P^{ }_2 = \K^{ }_1 + \K^{ }_2$. 
We introduce the usual Mandelstam invariants,
$s \equiv (\P^{ }_1 + \P^{ }_2)^2_{ }$, 
$t \equiv (\P^{ }_1 - \K^{ }_1)^2_{ }$, 
$u \equiv (\P^{ }_1 - \K^{ }_2)^2_{ }$.
The processes take place at very high temperatures, 
much above any known particles masses. Therefore
all particles can effectively be taken to be massless, whereby
\be 
 s + t + u = 0
 \;. \la{mandelstam_identity}
\ee
For massless fermions, sums over spins yield the standard expressions
$
 \sum_{\tau = \pm} u(\tau,\vec{p})\bar{u}(\tau,\vec{p})
 = 
 \sum_{\tau = \pm} v(\tau,\vec{p})\bar{v}(\tau,\vec{p})
 = \bsl{\P}
$.
For massless gauge bosons, a polarization sum yields
\be
 \sum_{\tau = \pm} \epsilon^\tau_{\mu}(\vec{p})\, 
                \epsilon^{\tau*}_{\nu}(\vec{p})
 \; \equiv \; 
 \mathbbm{P}^\rmii{T}_{\mu\nu}
 \;, 
 \la{P_T} 
\ee  
where $ \mathbbm{P}^\rmii{T}_{\mu\nu} $ is defined
like in \eq\nr{K_T}, but with four-momentum $\P$ rather than $\K$.
It projects onto two physical states in four dimensions, 
i.e.\ $|\mathbbm{P}^{\rmii{T}\mu}_\mu| = 2$. 

Inserting the Feynman rules to the amplitudes 
in \fig\ref{fig:graphs}, with the cubic graviton vertex being
particularly cumbersome~\cite{choi};  
constructing the amplitudes squared; 
and contracting with polarization or spin sums, as specified above, 
leads to rather lengthy expressions. We have found it practical to 
manipulate them with \textsc{FORM}~\cite{form}, 
\textsc{FeynArts}~\cite{feynarts}, 
\textsc{FormCalc}~\cite{formcalc},
and/or \textsc{FeynCalc}~\cite{feyncalc}.
In the course of these computations, we have verified the 
gauge independence of the contribution originating from massless
gauge bosons, notably that the last term of \eq\nr{K_T}
(with $\K\to\P$), 
containing $\bar\P$, does not contribute.

The matrix elements squared are best tabulated in a form where we sum not
only over all spins and polarization states, both of initial and final-state
particles, but also over particles and antiparticles. For a complex scalar 
field $\phi$, the processes from \fig\ref{fig:graphs}(a) 
then yield~\cite{gss}
\be
 \sum_\rmi{all} |\mathcal{M}^{ }_{\phi\phi^*_{ }\to h h}|^2_{ } + 
 \sum_\rmi{all} |\mathcal{M}^{ }_{\phi^*_{ }\phi\to h h}|^2_{ } 
 \; = \; 
 \frac{\kappa^4_{ }}{4} \frac{t^2_{ } u^2_{ }}{s^2_{ }} 
 \; \equiv \; 
 \frac{\kappa^4_{ }}{4}\,
 \Phi^{ }_\rmi{scalar} 
 \;. \la{M_scalar}
\ee
In the fermionic case, we take a Dirac fermion of definite 
chirality (i.e.\ a Weyl fermion) as a building block, with two 
physical degrees of freedom. However, as we sum over fermions
and antifermions, we effectively have four degrees of freedom
for each fermionic external leg, and therefore each of them can effectively 
be represented as a usual Dirac fermion $\psi$. The processes
in \fig\ref{fig:graphs}(b) then lead to 
\be
 \sum_\rmi{all} |\mathcal{M}^{ }_{\psi\bar\psi\to h h}|^2_{ }
 \; = \; 
 \frac{\kappa^4_{ }}{4}
 \biggl( - \frac{t^2_{ } u^2_{ }}{s^2_{ }} + \frac{t u}{2} \biggr)
 \; \equiv \; 
 \frac{\kappa^4_{ }}{4}\,
 \Phi^{ }_\rmi{fermion} 
 \;. \la{M_fermion}
\ee
Finally, for a U(1) gauge boson $g$, the processes 
in \fig\ref{fig:graphs}(c) produce 
\be
 \sum_\rmi{all} |\mathcal{M}^{ }_{gg\to h h}|^2_{ }
 \; = \; 
 \frac{\kappa^4_{ }}{4}
 \biggl(\, \frac{t^2_{ } u^2_{ }}{s^2_{ }} - 2\, t u 
       + \frac{s^2_{ }}{2} \,\biggr)
 \; \equiv \; 
 \frac{\kappa^4_{ }}{4}\,
 \Phi^{ }_\rmi{gauge} 
 \;. \la{M_gauge}
\ee

We note that, up to overall normalization, 
\eqs\nr{M_scalar} and \nr{M_fermion} can be extracted as the massless
limits of differential cross sections given in ref.~\cite{holstein2}. 
However, this is not the case for \eq\nr{M_gauge}, as the assignment of 
a mass to vector fields is ambiguous and breaks gauge invariance.
If masses are given to gauge fields, 
then the physical origin of the longitudinal polarization states 
should be specified
(e.g.\ through the Higgs mechanism), 
in order to obtain a gauge independent expression. 

\vspace*{3mm}

Given the matrix elements squared, we can determine the thermally 
averaged production rate of gravitational radiation. As a tool for this, 
we introduce the polarization-averaged phase-space distribution of 
gravitons, $f^{ }_\gw$. The energy density carried by 
gravitational waves of physical momentum $k$ can now be expressed as 
\be
 {\rm d} e^{ }_\gw \; = \; 2 k f^{ }_\gw
 \frac{{\rm d}^3_{ }\vec{k}}{(2\pi)^3_{ }}
 \;, \la{rate}
\ee
where the factor 2 counts the physical polarization states. 
In a locally Minkowskian time, making use of the rotational 
invariance of the thermal ensemble, 
the production rate can subsequently be expressed as 
\be
 \frac{{\rm d} e^{ }_\igw}{{\rm d}t\,{\rm d}\ln k}
 \; = \; 
 \frac{k^4 \dot{f}^{ }_\igw}{\pi^2_{ }}
 \;. \la{rate2}
\ee

The time variation of the phase-space distribution can 
be extracted from a Boltzmann equation. Assuming the particle content
of the Standard Model
(with three generations of charged leptons $\ell$, neutrinos $\nu$, 
and up and down-type quarks $u,d$, each with left and
right chiralities L and R), this takes at $\rmO(\kappa^4_{ })$ the form 
\ba
 \dot{f}^{ }_\gw & \stackrel{\rmO(\kappa^4_{ })}{\supset} & 
 \frac{1}{8k}
 \int \! {\rm d}\Omega^{ }_{2\to2}\,
 \frac{\kappa^4_{ }}{4}\,
 \biggl\{\,
 \nB^{ }(p^{ }_1)
 \nB^{ }(p^{ }_2)
 \,
 \biggl[\,
      \underbrace{2}_{\rm SU^{ }_L(2)} 
      \Phi^{ }_\rmi{scalar}
 \,\biggr]
 \nn
 & + & 
 \nF^{ }(p^{ }_1)
 \nF^{ }(p^{ }_2)
 \,
 \biggl[ 
      3\, \bigl(\, 
                  \underbrace{3}_{\ell^{ }_\rmiii{L},\ell^{ }_\rmiii{R},
                                  \nu^{ }_\rmiii{L}} 
                + \underbrace{1}_{\nu^{ }_\rmiii{R}} 
                + \underbrace{4\Nc^{ }}_{u^{ }_\rmiii{L},u^{ }_\rmiii{R},
                                         d^{ }_\rmiii{L},d^{ }_\rmiii{R}}
           \,\bigr)\,
      \Phi^{ }_\rmi{fermion}
 \biggr]
 \nn
 & + & 
 \nB^{ }(p^{ }_1)
 \nB^{ }(p^{ }_2)
 \,
 \biggl[
      \bigl(\,  
                  \underbrace{1}_{\rm U^{ }_Y(1)} 
                + \underbrace{3}_{\rm SU^{ }_L(2)} 
                + \underbrace{\;\Nc^2 - 1\;}_{\rm SU^{ }_c(\Nc^{ })} 
      \,\bigr)\,
      \Phi^{ }_\rmi{gauge}
 \biggr]
 \,\biggr\} 
 \;, \la{boltzmann}
\ea
where $p^{ }_i \equiv |\vec{p}^{ }_i|$, 
$\Nc^{ } = 3$, 
and the Bose and Fermi distributions have been introduced as 
$
 \nB^{ }(\epsilon) \; \equiv \; {1}/({e^{\epsilon/T}_{ } - 1})
$
and 
$
 \nF^{ }(\epsilon) \; \equiv \; {1}/({e^{\epsilon/T}_{ } + 1})
$, 
respectively. 
The prefactor $1/(8k)$ is a combination of the standard $1/(2k)$
associated with phase space, a factor $1/2$ for cancelling the sum 
over graviton polarizations that we had included in 
$\sum |\mathcal{M}|^2_{ }$, as well as $1/2$ for cancelling the overcounting
of identical particles or the redundant sum over particles and antiparticles
introduced above. 
Furthermore, we note that \eq\nr{boltzmann} includes only the gain terms, 
no loss terms, because the graviton phase-space density is much below the
equilibrium value, 
$
 f^{ }_\gw(k) \ll \nB^{ }(k)
$. 
For the same reason, 
there are no Bose enhancement factors
for the final-state gravitons in the gain terms. 

Let us remark that right-handed neutrinos have been included as degrees
of freedom in \eq\nr{boltzmann}. However, they only interact via 
Yukawa couplings, whose magnitudes are unknown. If they are small, 
right-handed neutrinos might not equilibrate fast enough to be part 
of the thermal ensemble, and should be omitted as initial states. 
We will illustrate the numerical difference between including and not
including right-handed neutrinos in \fig\ref{fig:spectrum}.
It is perhaps appropriate to mention that the 
equilibration rates of many other particles have also been discussed in 
the literature, however this typically concerns 
particle asymmetry changing reactions
relevant for chemical equilibration (cf.,\ e.g.,\ ref.~\cite{db}), 
whereas kinetic equilibrium should be efficiently maintained
by gauge interactions.

The phase-space average in \eq\nr{boltzmann} 
can be conveniently implemented by taking
the $s$-channel momentum transfer as the integration variable. For this, 
we write $\P^{ }_{i} = ({p}^{ }_i,\vec{p}^{ }_i)$
and $\K^{ }_{i} = ({k}^{ }_i,\vec{k}^{ }_i)$
with $\K \equiv \K^{ }_2$. 
Then, 
\ba
 \int \! {\rm d}\Omega^{ }_{2\leftrightarrow 2}
 & \equiv & 
 \int \! \frac{{\rm d}^3\vec{p}_1^{ }
             \,{\rm d}^3\vec{p}_2^{ } 
             \,{\rm d}^3\vec{k}_1^{ }
             \, {\rm d}^4\mathcal{Q} }
         {           (2 {p}^{ }_1)
                   \,(2 {p}^{ }_2)
                   \,(2 {k}^{ }_1)
           \, (2\pi)^9  }
  \, (2\pi)^4\, 
 \delta^{(4)}_{ }
                 \bigl(\P^{ }_1 + \P^{ }_2 - \mathcal{Q}\bigr)\,
 \delta^{(4)}_{ }
                 \bigl(\mathcal{Q}  - \K^{ }_1 - \K \bigr)
 \nn 
 & = & \!\!
 \int\! \frac{   {\rm d}^3\vec{p}_2^{ }
              \, {\rm d} q^{ }_0
              \, {\rm d}^3\vec{q} }
             {   |\vec{q-p}^{ }_2|
              \, {p}^{ }_2 
              \, |\vec{q-k}|
             }
 \, 
 \frac{ 
 \delta\bigl(   |\vec{q-p}^{ }_2|
              + {p}^{ }_2 
              - q^{ }_0
            \bigr) 
 \, 
 \delta\bigl(   q^{ }_0  
             -  |\vec{q-k}|
             -  k
       \bigr)
 }{8(2\pi)^5}
 \;, \hspace*{6mm} 
 \la{dPhi22}
\ea
where integrals were carried out over $\vec{p}^{ }_1$ and $\vec{k}^{ }_1$.
The constraints fix two angles as 
\be
 \cos\theta^{ }_{\vec{q},\vec{p}^{ }_2} \; = \; 
 \frac{q^2 + q^{ }_0 (2 p^{ }_2 - q^{ }_0)}{2 q p^{ }_2}
 \;, \quad
 \cos\theta^{ }_{\vec{q},\vec{k}} \; = \;
 \frac{q^2 + q^{ }_0 (2k - q^{ }_0)}{2 q k}
 \;. \la{angles1}
\ee
The remaining angle can be expressed in terms of an azimuthal variable, 
$\varphi$, as 
\be
 \cos\theta^{ }_{\vec{k},\vec{p}^{ }_2} \; = \; 
 \cos\theta^{ }_{\vec{q},\vec{k}} 
 \cos\theta^{ }_{\vec{q},\vec{p}^{ }_2} 
 + 
 \sin\theta^{ }_{\vec{q},\vec{k}} 
 \sin\theta^{ }_{\vec{q},\vec{p}^{ }_2} 
 \cos\varphi
 \;. \la{angles2}
\ee
Carrying out the angular integrals in order to remove the Dirac-$\delta$'s, 
we then obtain
\be
 \int \! {\rm d}\Omega^{ }_{2\leftrightarrow 2}
 \; = \; 
 \frac{1}{(4\pi)^3_{ }k}
 \int_k^\infty \! {\rm d}q^{ }_0 
 \int_{|2k - q^{ }_0|}^{q^{}_0} \! {\rm d}q 
 \int_{\qm^{ }}^{\qp^{ }} \! {\rm d}p^{ }_2 
 \int_{-\pi}^{\pi} \! \frac{{\rm d}\varphi}{2\pi}
 \;, \qquad
 q^{ }_{\pm} \; \equiv \; \frac{q^{ }_0 \pm q}{2}
 \;. \la{measure}
\ee
Subsequently, 
the integrals over $p^{ }_2$ and $\varphi$ can be performed analytically, 
as detailed in appendix~A. We represent the result as 
\be
 \int_{\qm^{ }}^{\qp^{ }} \! {\rm d}p^{ }_2 
 \int_{-\pi}^{\pi} \! \frac{{\rm d}\varphi}{2\pi}
 \,\nB^{ }(\underbrace{q^{ }_0 - p^{ }_2}_{p^{ }_1}) \nB^{ }(p^{ }_2) 
 \, \Phi^{ }_\rmi{scalar} 
 \; \equiv \; 
 \nB^{ }(q^{ }_0) 
 \, (q_0^2 - q^2_{ })^2_{ }
 \, \Theta^{ }_\rmi{scalar}
 \;, \la{Theta_def}
\ee
and correspondingly for 
$
 \Theta^{ }_\rmi{fermion}
$
and 
$
 \Theta^{ }_\rmi{gauge}
$. 
The expressions originating from 
\eqs\nr{M_scalar}, \nr{M_fermion} and \nr{M_gauge}
are given in 
\eqs\nr{Theta_scalar}, \nr{Theta_fermion} and \nr{Theta_gauge}, 
respectively. 

The physical production rate can now be assembled from 
\eqs\nr{rate2}, \nr{boltzmann}, \nr{measure}, and \nr{Theta_def}. 
It entails a 2-dimensional integral, 
\ba
 \frac{{\rm d} e^{ }_\igw}{{\rm d}t\,{\rm d}\ln k}
 & \supset &  
 \frac{\kappa^4_{ } k^2_{ }}{2(4\pi)^5_{ }}
 \int_k^\infty \! {\rm d}q^{ }_0 
 \int_{|2k - q^{ }_0|}^{q^{}_0} \! {\rm d}q 
 \, \nB^{ }(q^{ }_0) 
 \, (q_0^2 - q^2_{ })^2_{ }
 \nn[3mm] 
 & \times & 
 \bigl\{\,
      2\,
      \Theta^{ }_\rmi{scalar}
      + 
      12\, 
      (1 + \Nc^{ })\, 
      \Theta^{ }_\rmi{fermion} 
      +       
      (3 + \Nc^2) \, 
      \Theta^{ }_\rmi{gauge}
 \,\bigr\} 
 \;, \la{final}
\ea
which is 
rapidly convergent and 
readily evaluated numerically.\footnote{%
 In this respect the rate at $\rmO(\kappa^4_{ })$ differs from that
 at $\rmO(\alpha \kappa^2_{ })$: the latter is logarithmically 
 IR divergent in naive perturbation theory,  
 and requires Hard Thermal Loop resummation~\cite{htl3,htl4},
 in order to obtain a finite result~\cite{gravity_lo}.
 After the resummation, it contains a logarithmically 
 enhanced term, of 
 $\rmO(\alpha\kappa^2_{ }\ln(1/\alpha))$~\cite{gravity_qualitative}.
 } 

%
\section{Numerical results}
\la{se:plots}

%
\begin{figure}[t]

\centerline{%
     \epsfysize=7.6cm\epsfbox{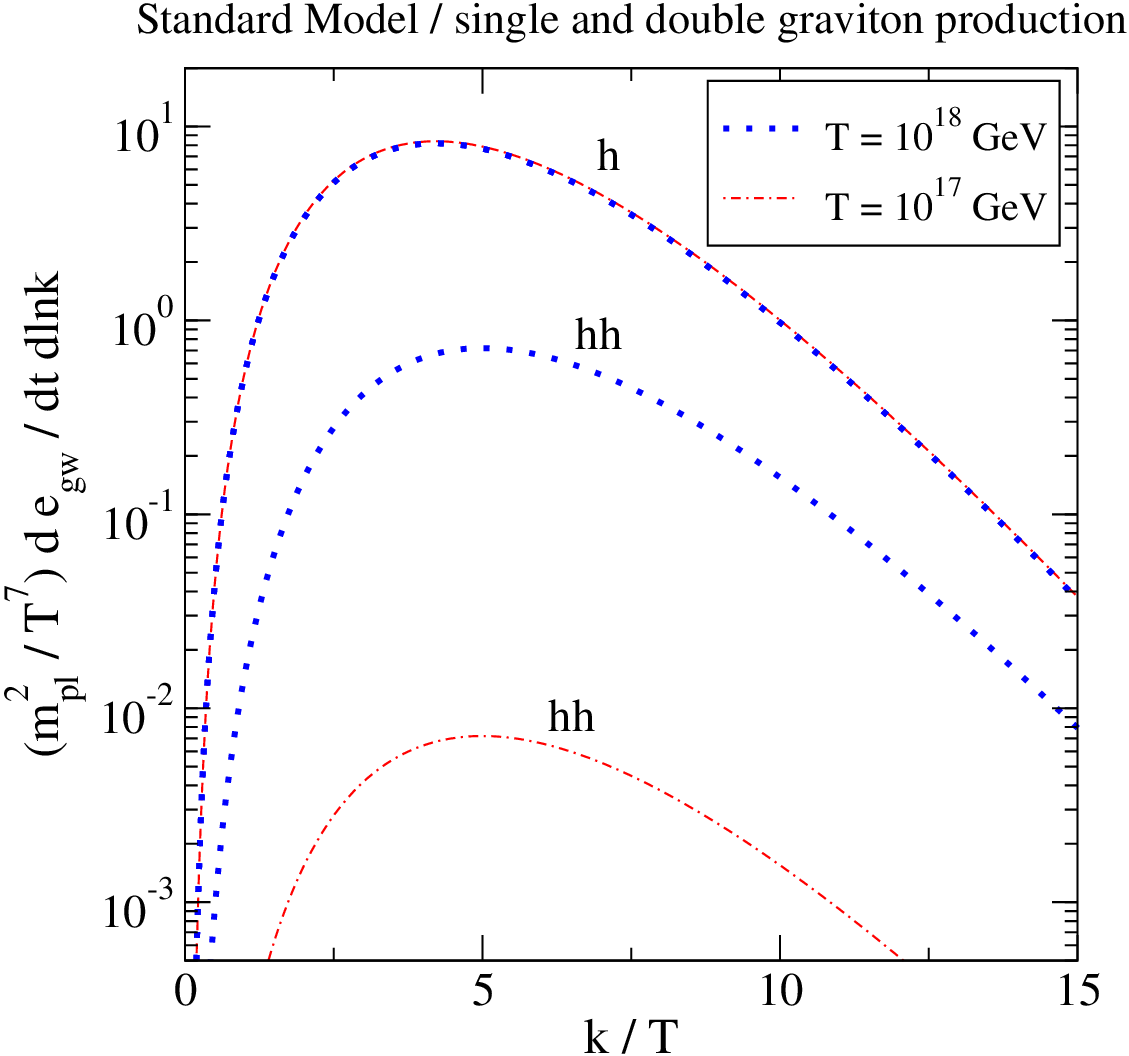}
  ~~~\epsfysize=7.8cm\epsfbox{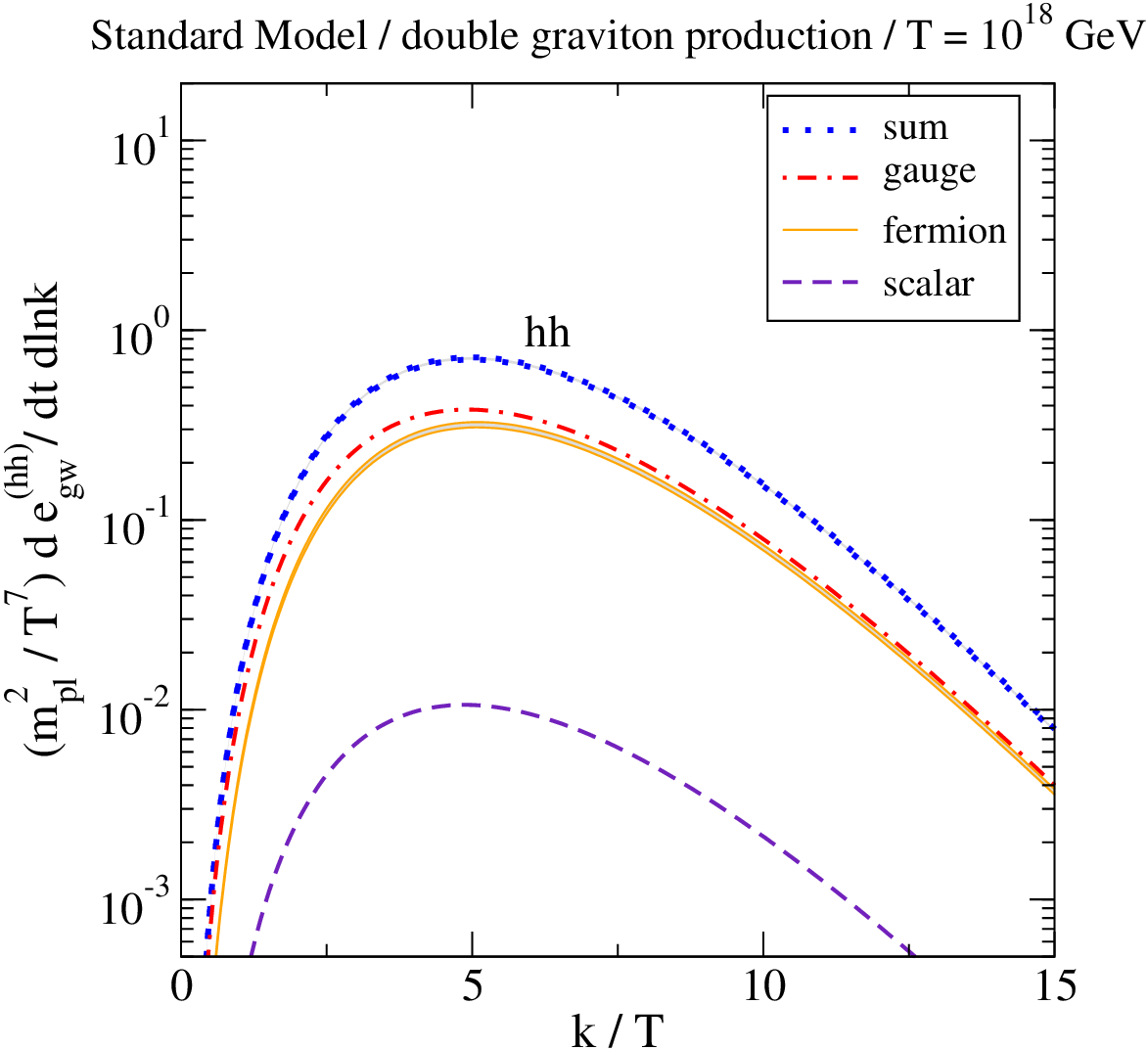}
}

\vspace*{1mm}

\caption[a]{\small 
 Left: the single~(h)~\cite{gravity_lo}
 and double (hh)
 graviton production rates from
 a Standard Model plasma.
 The slow evolution of the single-graviton rate  
 with the temperature is due to a logarithmic 
 running of the couplings, 
 whereas the double-graviton 
 rate scales
 with the temperature as $T^2_{ } / \mpl^2$
 in the units chosen. 
 At $T = 10^{18}_{ }$~GeV, the double-graviton rate is still 
 more than an order of magnitude smaller than the single-graviton rate, 
 but it would become dominant at $T > 4 \times 10^{18}_{ }$~GeV.  
 Right: different contributions to 
 the double-graviton rate from 
 \eq\nr{final} at $T = 10^{18}_{ }$~GeV. 
 In the fermionic part and the sum,
 the narrow band indicates the effect of 
 omitting right-handed neutrinos. 
} 

\la{fig:spectrum}
\end{figure}
%

%
\begin{figure}[t]

\centerline{%
     \epsfysize=7.6cm\epsfbox{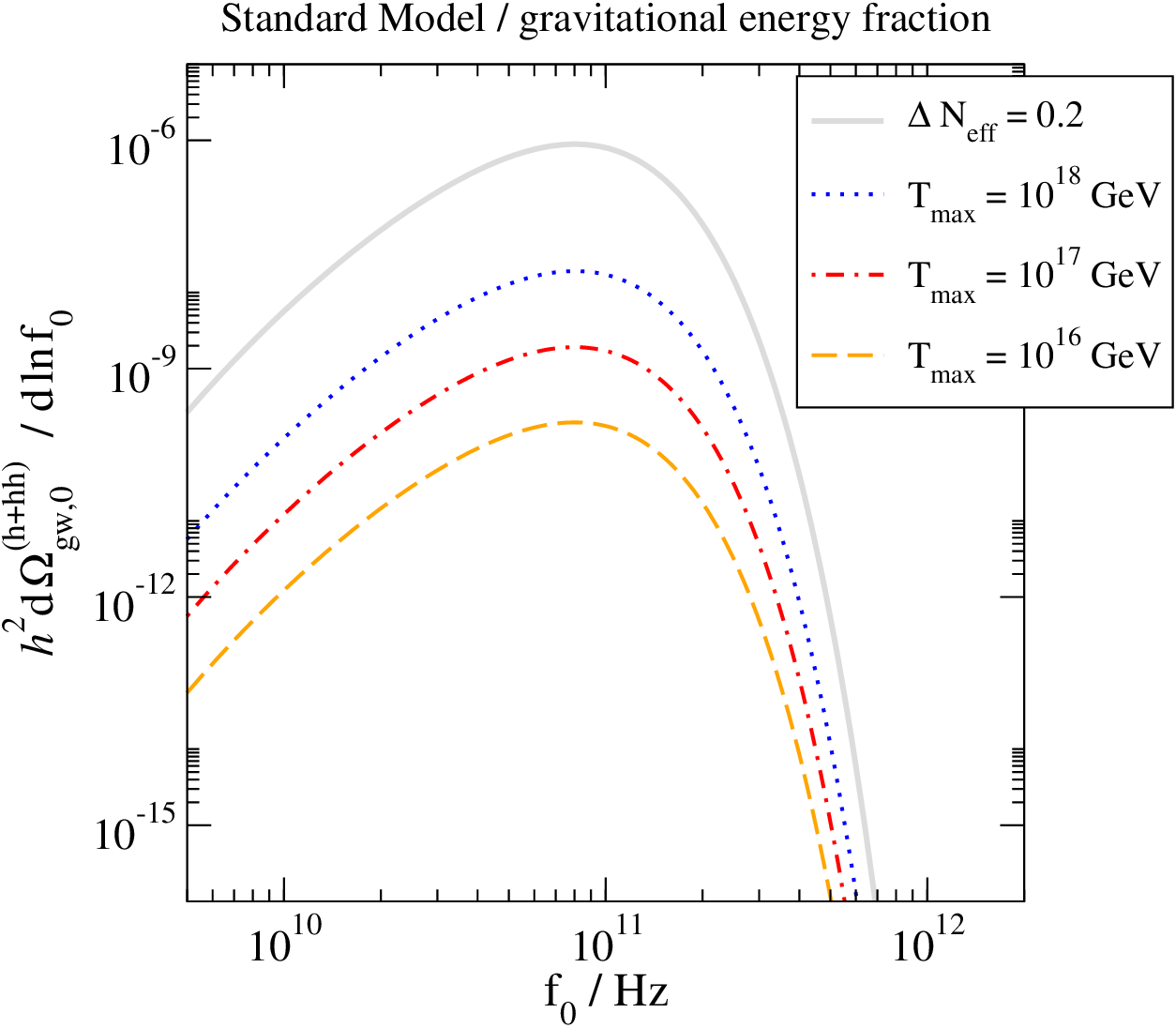}
}

\vspace*{1mm}

\caption[a]{\small 
 The time-integrated contribution of single and 
 double-graviton production to the differential
 gravitational energy density today, 
 $h^2 {\rm d}\Omega^{ }_\rmii{gw,0} / {\rm d}\ln \fnow^{ }$,
 as a function of the current-day frequency, $f^{ }_0$, 
 and various maximal temperatures, $\Tmax^{ }$.
 We have assumed radiation-dominated
 expansion at $T < \Tmax^{ }$~\cite{eos15}, 
 and denoted by $h$ the reduced Hubble rate. 
 The results are compared with a curve of the same
 shape (grey line), 
 which after integration over $\fnow^{ }$ 
 yields 
 $
 h^2_{ }\Omega^{ }_\rmii{gw,0} \equiv 5.62\times 10^{-6}_{ } 
 \Delta N^{ }_\rmii{eff}
 $ (cf.,\ e.g.,\ ref.~\cite{cf}). 
 Here we have set $\Delta N^{ }_\rmii{eff} = 0.2$, 
 which is the maximal observationally allowed 
 contribution to the effective number of massless
 neutrino species~\cite{planck6}. 
} 

\la{fig:Omega_gw}
\end{figure}
%

A numerical illustration of \eq\nr{final} is given 
in \fig\ref{fig:spectrum}. We have plotted the rate 
in units of $T^7_{ } / \mpl^2$, whereby it scales with 
the temperature as $T^2_{ } / \mpl^2$. The reason for this choice
is that the result can then be conveniently compared with
the rate originating from 
single-graviton production~\cite{gravity_lo}, 
which is practically constant in these units. 
For both results we have assumed the Standard Model matter content. 
Additional particle species could be included in 
a fairly straightforward 
manner~(cf.,\ e.g.,\ refs.~\cite{gravity_bsm,reheat,susy,kin,brems,hidden}), 
but we do not expect them to change the overall pattern, even though
they could have a quantitative influence.

The main observation from \fig\ref{fig:spectrum} is that
in the domain of the peak frequency, 
the double-graviton production rate is below the single-graviton 
production rate at $T < 4 \times 10^{18}_{ }$~GeV. In order to 
put this in context, we recall that 
the inflationary 
constraint $H^{ }_\rmi{max} < 10^{-5}_{ }\mpl^{ }$~\cite{planck10}, 
combined with the assumption of instantaneous reheating, 
suggests $\Tmax^{ } < 10^{16}_{ }$~GeV.
Therefore, within standard inflationary scenarios, 
the thermal part of the gravitational wave production rate can 
always be estimated from the single-graviton contribution. 

If we ignore the standard inflationary paradigm, 
another constraint on gravitational waves 
originates from $N^{ }_\rmi{eff}$, parametrizing the  
energy density residing in light degrees of freedom
 at the time of primordial nucleosynthesis
($
 \Delta e^{ }_{\gw,\bbn} 
 \equiv 
 \Delta N^{ }_\rmi{eff}\,
 ({7}/{8}) ({4}/{11})^{4/3}_{ } e^{ }_{\gamma,\bbn}
$). 
In order to obtain the gravitational contribution, we integrate 
the evolution of the phase space distribution, 
appearing in \eq\nr{rate2}, 
over the cosmological history, 
along a comoving trajectory~\cite{gravity_qualitative}.
The time integral can be converted into an integral 
over temperature, from the current $\Tnow^{ }$ 
up to a maximal one, $\Tmax^{ }$, yielding 
\be
 \frac{{\rm d} e^{ }_\rmiii{gw,0}}{{\rm d}\ln k(t^{ }_\rmiii{0})}
 = - 
 \int_{T^{ }_\rmiii{0}}^{T^{ }_\rmiii{max}}
 \! {\rm d} T \, 
 \frac{{\rm d}t}{{\rm d}T}\,
 \frac{a^4_{ }(T)}{a^4_{ }(T^{ }_\rmiii{0})}\,
 \biggl[\,
 \frac{{\rm d} e^{ }_\igw}{{\rm d}t\,{\rm d}\ln k}
 \,\biggr]^{ }_{k = k(t^{ }_\rmiii{0}) 
 \frac{a^{ }_{ }(T^{ }_\rmiii{0})}{a^{ }_{ }(T)}
 }
 \;, 
\ee
where $\tnow^{ }$ is the current time, 
$a$ is the cosmological scale factor, and
the relationship $t\leftrightarrow T$ is to be determined
from the Friedmann equations. 
Subsequently, normalizing with the current critical energy density, 
$
 e^{ }_\rmi{crit} = 
 h^2_{ } e^{ }_{\gamma,\rmii{0}} /  
 2.473 \times 10^{-5}_{ } / 
 (T^{ }_\rmii{0} / 2.7255~\mbox{K})^4_{ } 
$,
where 
$ e^{ }_{\gamma,\rmii{0}} $
is the current photon energy density
and $h$ is the reduced Hubble rate, 
and converting $k(t^{ }_\rmii{0})$ into 
the current frequency, $\fnow^{ }$,
yields the gravitational energy fraction, 
$
 h^2 {\rm d} \Omega^{ }_\rmii{gw,0} / {\rm d}\ln \fnow^{ }
$.\footnote{%
 In the literature, $ \Omega^{ }_\rmii{gw,0} $ sometimes denotes
 the differential spectrum, but here we need a separate notation
 for the differential 
 ($ {\rm d} \Omega^{ }_\rmii{gw,0} / {\rm d}\ln \fnow^{ } $) 
 and integrated spectra
 ($ \Omega^{ }_\rmii{gw,0} $). 
 }
The result is shown in \fig\ref{fig:Omega_gw}. 
Once we integrate over $\fnow^{ }$, 
obtaining 
$
 h^2 \Omega^{ }_\rmii{gw,0} 
$,
the total gravitational
energy density contributes to $N^{ }_\rmii{eff}$ as 
$
 \Delta N^{ }_\rmii{eff} \approx 
 h^2_{ }\Omega^{ }_\rmii{gw,0} /  5.62\times 10^{-6}_{ } 
$
(cf.,\ e.g., ref.~\cite{cf}).

Given that our result for 
$
 h^2_{ }\Omega^{ }_\rmii{gw,0}
$
depends on $\Tmax^{ }$, 
we now get a relationship between $\Tmax^{ }$
and $\Delta N^{ }_\rmi{eff}$. This can be compared
with existing theoretical and experimental knowledge
about $\Delta N^{ }_\rmi{eff}$. In particular, 
the current estimated uncertainty of the 
Standard Model prediction, 
$\Delta N^{ }_\rmi{eff} \le 0.001$, corresponds to 
$T^{ }_\rmi{max} \le 2\times 10^{17}$~GeV~\cite{gravity_lo}.
At $\Tmax^{ } = 4 \times 10^{18}_{ }$~GeV, when the 
double-graviton rate overtakes the single-graviton rate, 
$\Delta N^{ }_\rmi{eff} \approx 0.02$.\footnote{%
 As mentioned in \se\ref{se:intro}, 
 the single-graviton production rate involves $\sim \alpha/ \mpl^2$, 
 but the age of the universe scales with $\sim\mpl^{ }$,
 whereby the integrated 
 contribution to $N^{ }_\rmii{eff}$ 
 behaves as $\sim \alpha \, \Tmax^{ }/\mpl^{ }$.
 } 
This is still below the current experimental limit, 
$\Delta N^{ }_\rmi{eff} < 0.2$~\cite{planck6}. 
For illustration, 
a spectrum that would yield
$\Delta N^{ }_\rmi{eff} = 0.2$
is also shown in \fig\ref{fig:Omega_gw}.

%
\section{Conclusions and outlook}
\la{se:concl}

In this work we have computed the double-graviton production rate 
from a thermal Standard Model plasma present during an early 
radiation-dominated epoch. Previous calculations had only considered
scalar field contributions to the double-graviton rate~\cite{gss}. 
For momenta $k\sim \pi T$, the double-graviton rate scales as
$
 {{\rm d} e^{ }_\igw} / 
 [ {{\rm d}t\,{\rm d}\hspace*{-0.3mm}\ln k} ]
 \sim T^{\hspace*{0.2mm}9}_{ }/ \mpl^4
$. 
This can be compared with the single-graviton production 
rate~\cite{gravity_lo}, which is of order 
$
 {{\rm d} e^{ }_\igw} / 
 [ {{\rm d}t\,{\rm d}\hspace*{-0.3mm}\ln k} ]
 \sim \alpha\, T^{\hspace*{0.2mm}7}_{ }/ \mpl^2
$,
where $\alpha$ is a fine-structure constant.

Our result for the double-graviton production rate is given
in \eq\nr{final} and illustrated numerically 
in \fig\ref{fig:spectrum}. 
A comparison with the single-graviton production rate
shows that the 
$\rmO( T^{\hspace*{0.2mm}9}_{ }/ \mpl^4 )$ contribution overtakes the 
$\rmO( \alpha\, T^{\hspace*{0.2mm}7}_{ }/ \mpl^2 )$ one at 
$T \sim 4 \times 10^{18}_{ }$~GeV.
This temperature is much higher than expected to be reached after
standard inflationary scenarios. Therefore, 
single-graviton production is normally the dominant source.
If one considers BSM scenarios associated with a non-renormalizable 
coupling, $\sim 1/\Lambda^2_{ }$ (cf.,\ e.g.,\ ref.~\cite{reheat}), 
then we should effectively replace $\alpha\to T^2/\Lambda^2$
in the single-graviton rate. In these cases, if $\Lambda\sim \mpl^{ }$, 
the single and double-graviton contributions could be 
of comparable magnitudes.

In order to obtain our results, we needed to evaluate the
matrix elements squared corresponding to the Feynman diagrams
in \fig\ref{fig:graphs}, producing the results given 
in \eqs\nr{M_scalar}--\nr{M_gauge}. 
Similar matrix elements squared have
been considered in previous literature, 
either out of theoretical interest 
(cf.,\ e.g.,\ ref.~\cite{holstein2}), 
or in view of the scattering 
of gravitational waves on matter
(cf.,\ e.g., ref.~\cite{astro}).
As we are concerned with very high 
temperatures, the Standard Model particles can be considered
as massless for the purposes of our analysis. 
We noted 
that the literature results had assigned masses
to gauge fields in a way which did not permit
to recover the massless limit of \eq\nr{M_gauge}. 

Let us elaborate briefly on the more general conceptual
implications of our results. As explained below \eq\nr{kappa_def},
the gravitational wave production rate within General Relativity
is a double series, in $\alpha$ and $T^2/\mpl^2$. Normally, to 
estimate the convergence of the series in $T^2/\mpl^2$, we would compare
terms with a same power of $\alpha$. However, the term of 
$\rmO(\alpha^0_{ }T^2/\mpl^2)$ is absent, because it is 
kinematically forbidden, and the term of $\rmO(\alpha^1_{ }T^4/\mpl^4)$
is unknown, as it is of next-to-leading order, either
in $\alpha$ 
with respect to the known contribution of $\rmO(\alpha^0_{ }T^4 / \mpl^4)$, 
or in $T^2_{ }/ \mpl^2$ with respect to the known contribution of 
$\rmO(\alpha^1_{ }T^2/\mpl^2)$. But we can still compare the known terms. 
Because of the absence of $\alpha$, the term of 
$\rmO(\alpha^0_{ }T^4 / \mpl^4)$ overtakes  the
$\rmO(\alpha^1_{ }T^2/\mpl^2)$ contribution {\em sooner} than 
the proper probe of $\rmO(\alpha^1_{ }T^4 / \mpl^4)$ would. 
Therefore, the criterion 
$T < 4 \times 10^{18}_{ }$~GeV
can be seen as a {\em conservative} (or sufficient) condition for 
the highest temperature at which General Relativity together
with the Standard Model can be 
employed as a self-consistent effective theory.  

Returning to the phenomenological side, 
we end by remarking that reaching a high temperature 
after inflation does not guarantee 
a substantial thermal gravitational wave background. 
The reason is that if the inflaton equilibration
rate is small compared with the Hubble rate, as may be
expected for an extremely weakly coupled field, then inflation is followed
by a period of matter domination (cf.,\ e.g.,\ ref.~\cite{matter}). 
Similarly to a phase transition produced 
signal (cf.,\ e.g.,\ ref.~\cite{uhf_pt}), 
the thermal background would be diluted during such a stage. 

%
\section*{Acknowledgements}

J.G.\ was supported by 
the Agence Nationale de la Recherche (France), under grant
ANR-22-CE31-0018 (AUTOTHERM),
J.S.-E.\ by the National Science
Foundation under cooperative agreement 202027 and by 
the RIKEN-Berkeley Center, 
and E.S.\ by
the European Union's Horizon Europe Research
and Innovation Program,
under the Marie Sk{\l}odowska-Curie
grant agreement no.\ 101109747.

%
\appendix
\renewcommand{\thesection}{\Alph{section}}
\renewcommand{\thesubsection}{\Alph{section}.\arabic{subsection}}
\renewcommand{\theequation}{\Alph{section}.\arabic{equation}}

%
\section{Azimuthal and radial momentum integrals}

We show here how two of the integrals in \eq\nr{measure} can be 
carried out analytically, in order to determine the factors $\Theta$
defined in \eq\nr{Theta_def} that 
enter our final result in \eq\nr{final}. 

Making use of the identities
\ba
 \nB^{ }(\epsilon^{ }_2) \nB^{ }(q^{ }_0 - \epsilon^{ }_2)
 & = &
 \nB^{ }(q^{ }_0) \, 
 \bigl[ 1 + \nB^{ }(\epsilon^{ }_2) 
          + \nB^{ }(q^{ }_0 - \epsilon^{ }_2) \bigr]
 \;, \la{identities1} \\ 
 \nF^{ }(\epsilon^{ }_2) \nF^{ }(q^{ }_0 - \epsilon^{ }_2)
 & = &
 \nB^{ }(q^{ }_0) \, 
 \bigl[ 1 - \nF^{ }(\epsilon^{ }_2) 
          - \nF^{ }(q^{ }_0 - \epsilon^{ }_2) \bigr]
 \;, \la{identities2}
\ea
the definition in \eq\nr{Theta_def} as well as the matrix
element squared in \eq\nr{M_scalar} imply that 
\be
 \Theta^{ }_\rmi{scalar}
 \; = \; 
 \int_{\qm^{ }}^{\qp^{ }}
 \! {\rm d} p^{ }_2 
 \int_{-\pi}^{\pi} \! \frac{{\rm d}\varphi}{2\pi} \,
 \biggl(\,
  \frac{t^2_{ }u^2_{ }}{s^4_{ }} 
 \,\biggr)
 \, 
 \bigl[
  1 + \nB^{ }(p^{ }_2) + \nB^{ }(q^{ }_0 - p^{ }_2) 
 \bigr]
 \;. \la{Theta_scalar_2}
\ee
Here $u = - s - t$, so that the integrand is a polynomial in $t/s$.
Inserting four-momenta, we get 
$t/s = 2 (\vec{k}\cdot\vec{p}^{ }_2 - k p^{ }_2)/s$, 
which after the use of  
\eqs\nr{angles1} and \nr{angles2}
can be expressed as 
\be
 \frac{t}{s} = 
 \frac{
   (2k - q^{ }_0)(2p^{ }_2 - q^{ }_0) - q^2_{ } 
  + \cos\varphi\, 
   \sqrt{q^2 - (2 k - q^{ }_0 )^2_{ }}
   \sqrt{q^2 - (2 p^{ }_2 - q^{ }_0 )^2_{ }}
 }{2q^2_{ }}
 \;. \la{tos}
\ee
The azimuthal average in \eq\nr{Theta_scalar_2} reduces to 
averages of $\cos^n_{ }\!\varphi$, trivially carried out. Left over 
are integrals over powers of $p^{ }_2$, weighted by a Bose or 
Fermi distribution. These can be expressed in terms of polylogarithms, 
which we define as 
\ba
 && 
 \lnb(\epsilon) \; \equiv \; \ln \Bigl( 1 - e^{-\epsilon/T} \Bigr)
 \;, \quad\; 
 \lnf(\epsilon) \; \equiv \; \ln \Bigl( 1 + e^{-\epsilon/T} \Bigr)
 \;, \la{logs} \\
 &&
 \lb{n}(\epsilon) \;\, \equiv \; \mbox{Li}^{ }_n \Bigl(e^{-\epsilon/T}\Bigr)
 \;, \quad \hspace*{5mm} 
 \lf{n}(\epsilon) \;\, \equiv \; \mbox{Li}^{ }_n \Bigl(-e^{-\epsilon/T}\Bigr)
 \;, \quad 
 n \ge 2
 \;. \la{polylogs}
\ea
Thereby 
\ba
 \Theta^{ }_\rmi{scalar}
 & = & 
 \frac{q}{30} 
 \; + \;  
 \frac{T [q^2_{ } - (2k-q^0_{ })^2_{ }]^2_{ }}{8 q^4_{ }}
 \bigl[\, \lnb(\qp^{ }) - \lnb(\qm^{ }) \,\bigr]
\nn[3mm]
 & - &  
 \frac{T^2_{ } [q^2_{ } - (2k-q^0_{ })^2_{ }]
               [q^2_{ } - 5 (2k-q^0_{ })^2_{ }] }{2 q^5_{ }}
 \bigl[\, \lb{2}(\qp^{ }) + \lb{2}(\qm^{ }) \,\bigr]
\nn[3mm]
 & - &  
 \frac{T^3_{ } [5 q^4_{ } - 42 q^2_{ } (2k-q^0_{ })^2_{ }
              + 45 (2k-q^0_{ })^4_{ }] }{2 q^6_{ }}
 \bigl[\, \lb{3}(\qp^{ }) - \lb{3}(\qm^{ }) \,\bigr]
\nn[3mm]
 & - &  
 \frac{3 T^4_{ } [3 q^4_{ } - 30 q^2_{ } (2k-q^0_{ })^2_{ }
              + 35 (2k-q^0_{ })^4_{ }] }{q^7_{ }}
 \bigl[\, \lb{4}(\qp^{ }) + \lb{4}(\qm^{ }) \,\bigr]
\nn[3mm]
 & - &  
 \frac{6 T^5_{ } [3 q^4_{ } - 30 q^2_{ } (2k-q^0_{ })^2_{ }
              + 35 (2k-q^0_{ })^4_{ }] }{q^8_{ }}
 \bigl[\, \lb{5}(\qp^{ }) - \lb{5}(\qm^{ }) \,\bigr]
 \;, \la{Theta_scalar}
\ea
\ba
 \Theta^{ }_\rmi{fermion}
 & = & 
 \int_{\qm^{ }}^{\qp^{ }}
 \! {\rm d} p^{ }_2 
 \int_{-\pi}^{\pi} \! \frac{{\rm d}\varphi}{2\pi} \,
 \biggl(\,
  -\frac{t^2_{ }u^2_{ }}{s^4_{ }} + \frac{tu}{2s^2_{  }} 
 \,\biggr)
 \, 
 \bigl[
  1 - \nF^{ }(p^{ }_2) - \nF^{ }(q^{ }_0 - p^{ }_2) 
 \bigr]
 \\[3mm]
 & = & 
 \frac{q}{20} 
 \; + \;  
 \frac{T [q^4_{ } - (2k-q^0_{ })^4_{ }] }{8 q^4_{ }}
 \bigl[\, \lnf(\qp^{ }) - \lnf(\qm^{ }) \,\bigr]
\nn[3mm]
 & + &  
 \frac{T^2_{ } [ - 3 q^2_{ } 
                 + 5 (2k-q^0_{ })^2_{ }] (2k-q^0_{ })^2_{ } }{2 q^5_{ }}
 \bigl[\, \lf{2}(\qp^{ }) + \lf{2}(\qm^{ }) \,\bigr]
\nn[3mm]
 & + &  
 \frac{3 T^3_{ } [q^4_{ } - 12 q^2_{ } (2k-q^0_{ })^2_{ }
              + 15 (2k-q^0_{ })^4_{ }] }{2 q^6_{ }}
 \bigl[\, \lf{3}(\qp^{ }) - \lf{3}(\qm^{ }) \,\bigr]
\nn[3mm]
 & + &  
 \frac{3 T^4_{ } [3 q^4_{ } - 30 q^2_{ } (2k-q^0_{ })^2_{ }
              + 35 (2k-q^0_{ })^4_{ }] }{q^7_{ }}
 \bigl[\, \lf{4}(\qp^{ }) + \lf{4}(\qm^{ }) \,\bigr]
\nn[3mm]
 & + &  
 \frac{6 T^5_{ } [3 q^4_{ } - 30 q^2_{ } (2k-q^0_{ })^2_{ }
              + 35 (2k-q^0_{ })^4_{ }] }{q^8_{ }}
 \bigl[\, \lf{5}(\qp^{ }) - \lf{5}(\qm^{ }) \,\bigr]
 \;, \la{Theta_fermion}
\ea
\ba
 \Theta^{ }_\rmi{gauge}
 & = & 
 \int_{\qm^{ }}^{\qp^{ }}
 \! {\rm d} p^{ }_2 
 \int_{-\pi}^{\pi} \! \frac{{\rm d}\varphi}{2\pi} \,
 \biggl(\,
  \frac{t^2_{ }u^2_{ }}{s^4_{ }} - \frac{2 t u}{s^2} + \frac{1}{2} 
 \,\biggr)
 \, 
 \bigl[
  1 + \nB^{ }(p^{ }_2) + \nB^{ }(q^{ }_0 - p^{ }_2) 
 \bigr]
 \\[3mm]
 & = & 
 \frac{q}{5} 
 \; + \;  
 \frac{T [q^4_{ } + 6 q^2 (2k-q^0_{ })^2_{ } 
                  + (2k-q^0_{ })^4_{ }] }{8 q^4_{ }}
 \bigl[\, \lnb(\qp^{ }) - \lnb(\qm^{ }) \,\bigr]
\nn[3mm]
 & + &  
 \frac{T^2_{ } [3 q^4_{ } - 6 q^2_{ } (2k-q^0_{ })^2_{ }
                - 5 (2k-q^0_{ })^4_{ }] }{2 q^5_{ }}
 \bigl[\, \lb{2}(\qp^{ }) + \lb{2}(\qm^{ }) \,\bigr]
\nn[3mm]
 & + &  
 \frac{3T^3_{ } [ q^4_{ } + 6 q^2_{ } (2k-q^0_{ })^2_{ }
              - 15 (2k-q^0_{ })^4_{ }] }{2 q^6_{ }}
 \bigl[\, \lb{3}(\qp^{ }) - \lb{3}(\qm^{ }) \,\bigr]
\nn[3mm]
 & - &  
 \frac{3 T^4_{ } [3 q^4_{ } - 30 q^2_{ } (2k-q^0_{ })^2_{ }
              + 35 (2k-q^0_{ })^4_{ }] }{q^7_{ }}
 \bigl[\, \lb{4}(\qp^{ }) + \lb{4}(\qm^{ }) \,\bigr]
\nn[3mm]
 & - &  
 \frac{6 T^5_{ } [3 q^4_{ } - 30 q^2_{ } (2k-q^0_{ })^2_{ }
              + 35 (2k-q^0_{ })^4_{ }] }{q^8_{ }}
 \bigl[\, \lb{5}(\qp^{ }) - \lb{5}(\qm^{ }) \,\bigr]
 \;. \la{Theta_gauge}
\ea

\small{
%

}

\end{document}